\documentclass[]{elsart}
\usepackage{graphicx}
\usepackage{epsfig}
\usepackage{amssymb}
\usepackage{dcolumn}
\usepackage{bm}

\begin{document}

\begin{frontmatter}

\title{Effects of stoichiometry, purity, etching and distilling on resistance of MgB$_{2}$ pellets and wire segments}

\author{R. A. Ribeiro}, \author{S. L. Bud'ko}, \author{C. Petrovic}, \author{P. C.
Canfield}

\address{Ames Laboratory and Department of Physics and Astronomy\\
Iowa State University, Ames, IA 50011 USA}

\begin{abstract}
We present a study of the effects of non-stoichiometry, boron
purity, wire diameter and post-synthesis treatment (etching and Mg
distilling) on the temperature dependent resistance and
resistivity of sintered MgB$_2$ pellets and wire segments. Whereas
the residual resistivity ratio ($RRR$) varies between $RRR$
$\approx 4$ to $RRR$ $\geq$ 20 for different boron purity, it is
only moderately affected by non-stoichiometry (from 20$\%$ Mg
deficiency to 20$\%$ Mg excess) and is apparently independent of
wire diameter and presence of Mg metal traces on the wire surface.
The obtained set of data indicates that $RRR$ values in excess of
20 and residual resistivities as low as $\rho_{0} \approx 0.4 \mu
\Omega cm$ are intrinsic material properties of high purity
MgB$_2$.
\end{abstract}

\end{frontmatter}

\section{Introduction}
\label{sec:intro}

Within weeks of the announcement of the discovery of
superconductivity in MgB$_{2}$ by Akamitsu and co-workers
\cite{sendai,akimitsu}, it was established that high purity, very
low residual resistivity samples of MgB$_{2}$ could be synthesized
by exposing boron powder or filaments to Mg vapor at temperatures
at or near $950^{\circ}C$ for as little as two hours
\cite{budkoPRL,finnemore01,canfield01}. Samples with residual
resistivity ratio [$RRR = R(300 K)/R(42 K)$] values in excess of
20 and residual resistivities as low as 0.4 $\mu \Omega cm$ were
synthesized by this method. Such a low resistivity in an
intermetallic compound with a superconducting critical
temperature, $T_{c}$, near $40 K$ was of profound physical, as
well as engineering, interest. The implications of this high $RRR$
and low $\rho_0$ ranged from large magneto-resistances (in
accordance with Kohler's rule) to questions of how a material with
such an apparently large electron-phonon coupling could have such
a small resistivity. On the applied side, a normal state
resistivity of 0.4 $\mu \Omega cm$ for temperatures just above
$T_{c}$ means that MgB$_{2}$ wires would be able to handle a
quench with much greater ease than, for example, Nb$_{3}$Sn wires
which have a $\rho_0$ that is over an order of magnitude larger
for $T \sim 20 K$ \cite{canfield01}.

Unfortunately other techniques of synthesizing MgB$_{2}$ have not
yet been able to achieve such high $RRR$ or low $\rho_0$ values
\cite{chen,lee,pradhan,kijoon,xue}. In some cases the authors of
these papers have concluded that the resistivity of their samples
must be the intrinsic resistivity and that higher $RRR$ values or
lower residual resistivity values must somehow be extrinsic. In
order to address these concerns and in order to shed some light on
how low resistivity samples can be grown we have studied the
effects of boron purity and magnesium stoichiometry on sintered
pellet samples. In addition we have studied the effects of
filament diameter and post synthesis etching and distilling on
MgB$_{2}$ wire segments. Based on these measurements we conclude
that the purity of the boron used to make the MgB$_{2}$ is a
dominant factor in determining the ultimate, low temperature,
normal state resistivity of the sample, and that $RRR$ values as
high as 20 and residual resistivities as low as 0.4 $\mu \Omega
cm$ are intrinsic materials properties of high purity MgB$_{2}$.

\section{Sample synthesis}
\label{sec:synthesis}

Samples of MgB$_{2}$ for this study were made in the form of
sintered pellets as well as wire segments. The sintered pellets
were made by sealing stoichiometric amounts of Mg and B into Ta
tubes and placing these tubes (sealed in quartz) into furnaces
heated to $950^{\circ}C$ for 3 hours and then quenched to room
temperature \cite{budkoPRL}. For the initial studies of boron
purity stoichiometric MgB$_{2}$ was synthesized and the quality of
the boron was varied. For the studies of magnesium stoichiometry
nominal stoichiometries that ranged from Mg$_{0.8}$$^{11}$B$_{2}$
to Mg$_{1.2}$$^{11}$B$_{2}$ were used and samples were synthesized
with isotopically 99.95\% enriched $^{11}$B. For all synthesizes
lump Mg of 99.9\% purity was used.

Wire segments of MgB$_{2}$ were made by sealing boron filaments
purchased from Textron \cite{textron} or Goodfellow
\cite{goodfellow} into a Ta tube with excess Mg, using a ratio of
approximately Mg$_{3}$B. After reacting the filaments for 68 hours
at $950^{\circ}C$ the Ta tubes were quenched to room temperature
and the wire segments were removed from the Ta reaction vessel.
Given that there can be some excess Mg on the surface of the wire
segments, some of the wire segments were etched in a solution of
5\% HCl and ethyl alcohol for times up to 5 minutes. This
treatment removes the surface Mg and leads to the surface of the
wire segments having the same appearance as the surface of the
stoichiometric sintered pellets: a slightly golden / grey color.
Another method was used to remove any potential surface Mg from
the MgB$_{2}$ wire segments: high temperature distillation of the
Mg. In order to achieve this a wire sample was place into a quartz
tube that was continuously pumped by a turbomolecular pump to a
pressure less than $10^{-5} Torr$. The evacuated tube was then
heated to $600^{\circ}C$ for 12 hours. This temperature and time
were chosen in part because attempts at higher temperature
distillation lead to a decomposition of the MgB$_{2}$ in the wire
itself.

A.C. electrical resistance measurements were made using Quantum
Design MPMS and PPMS units. Platinum wires for standard four-probe
configuration with connected to the sample with Epotek H20E silver
epoxy. LR 400 and LR 700 A.C. resistance bridges were used to
measure the resistivity when the MPMS units were used to provide
the temperature environment. Powder X-ray diffraction measurements
were made using a Cu $K_{\alpha}$ radiation in a Scintag
diffractometer and a Si standard was used for all runs. The Si
lines have been removed from the X-ray diffraction data, leading
to apparent gaps in the powder X-ray spectra.

\section{Effects of Boron Purity}
\label{sec:boron}

\begin{center}
\small Table 1: Boron form and purity (As provided by the seller).
\begin{tabular}{|c|c|c|c|} \hline
  \bf{Purity} & \bf{Form} & \bf{Source} & \bf{Main Impurities}  \\\hline
  90\%   & \begin{tabular}{c}
     Amorphous \\
    (325 mesh) \\
  \end{tabular} & Alfa Aesar & \begin{tabular}{c}
     Mg \\
    5\%\\
  \end{tabular}\\\hline
  95\%   &  \begin{tabular}{c}
     Amorphous \\
    ($<$ 5 mesh) \\
  \end{tabular} &Alfa Aesar  &\begin{tabular}{c}
     Mg \\
    1\%\\
  \end{tabular}\\\hline
  98\%   & \begin{tabular}{c}
     Crystalline  \\
    (325 mesh) \\
  \end{tabular} & Alfa Aesar  &\begin{tabular}{c}
     C \\
    0.55\%\\
  \end{tabular}\\\hline
 99.95\%& \begin{tabular}{c}
     Isotopically pure $^{11}$B \\
    Crystalline (325 mesh) \\
  \end{tabular} & Eagle-Picher &\begin{tabular}{c}
     Si\\
    0.04\%\\
  \end{tabular}\\\hline
 99.99\% & \begin{tabular}{c}
     Amorphous \\
    (325 mesh) \\
  \end{tabular} & Alfa Aesar  & \begin{tabular}{c}
     Metallic Impurities \\
    0.005\%\\
   \end{tabular}\\ \hline
\end{tabular}
\end{center}

Table 1 presents the source and purity information available for
each of the starting boron powders used. Figure 1 presents powder
X-ray diffraction spectra for three samples with varying nominal
boron purities: 90\% purity, 99.99\% purity, and the 99.95\%
purity, isotopically pure $^{11}$B. By comparing the two upper
panels to the bottom panel it can be seen that the strongest
MgB$_{2}$ lines are present in all three samples. The upper panel,
the data taken on the sample made from boron with only a nominal
90\% purity, also has a weak Mg and MgO powder lines   present.
This is not inconsistent with the fact that the primary impurity
in the 90\% boron is associated with Mg.

\begin{figure}[h]
\begin{center}
\includegraphics[angle=0,width=136mm]{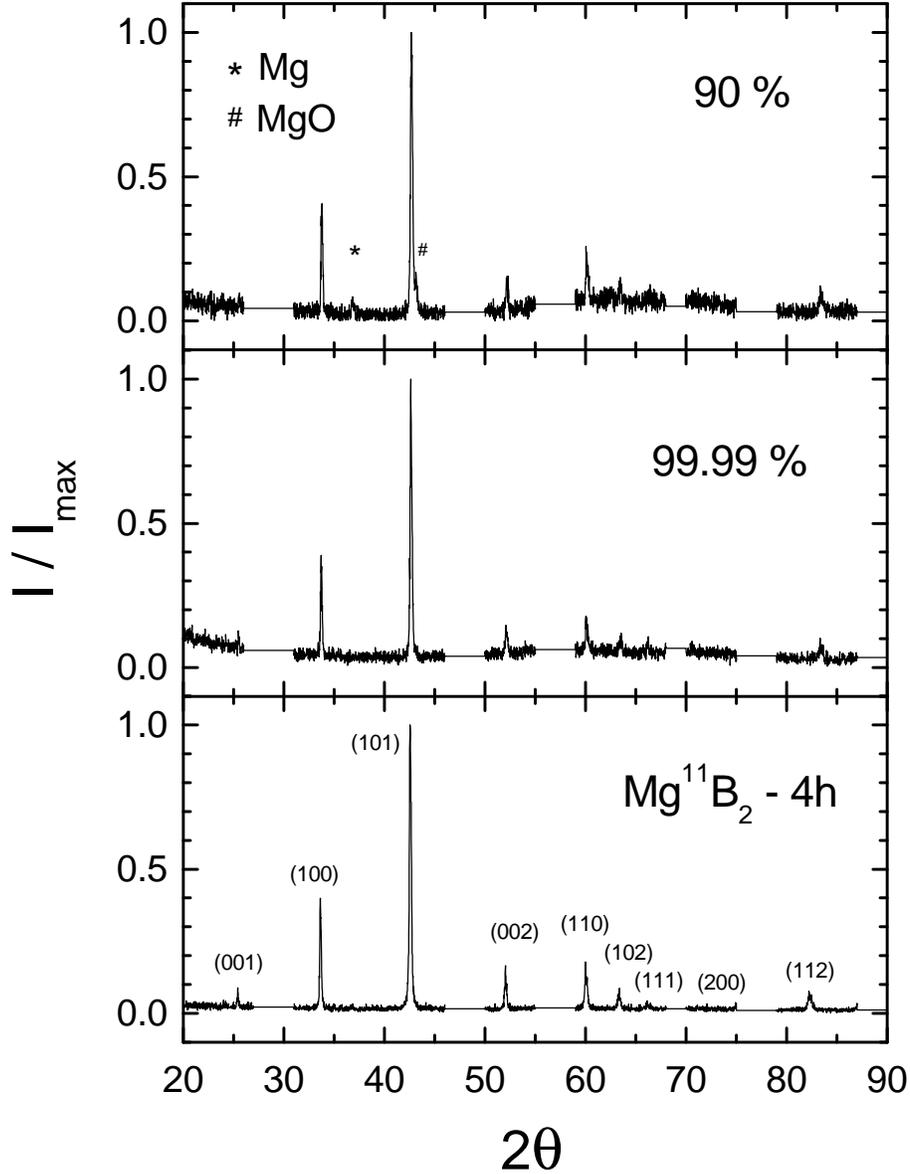}
\small\caption{\label{fig1} Powder x-ray (Cu $K_{\alpha}$
radiation) diffraction spectra of MgB$_{2}$ (with $h,k,l$) for 3
different qualities (a) pure natural boron 90\% ; (b)pure natural
boron 99.99\% and (c) isotopic $^{11}$B 99.95\%. Samples (a) and
(b) were synthesized for $3h/950^{\circ}C$, and sample (c) for
$4h/950^{\circ}C$ from\cite{cunningham}. The data gaps are due to
the removal of the Si peaks. The symbol $\star$ indicates a Mg
peak and $\#$ indicates a MgO peak.}
\end{center}
\end{figure}

Figure 2 displays the normalized resistance, $R(T)/R(300 K)$, of
MgB$_{2}$ pellets that were made using five different types of
boron powder. Each curve is the average of three resistivity
curves taken on different pieces broken off of each pellet. Figure
2 demonstrates that $RRR$ values can range from as low as 4 to as
high as 20 depending upon what source of boron is used. Among the
natural boron samples examined there is a steady increase in $RRR$
as the purity of the source boron is improved. The MgB$_{2}$
synthesized from the isotopically pure boron appears to have the
best $RRR$, although it's nominal purity is somewhat less than
that of the 99.99\% pure natural boron, but those skilled in the
art will realize that claims of purity from different companies
can vary dramatically. In addition, it is very likely that the
isotopically pure boron was prepared in a somewhat different
manner from the other boron powders used (very likely using a
boron fluoride or boric acid or any of its complexes as an
intermediate phase in order to achieve isotopic separation). The
primary point that figure 2 establishes is that the purity of the
boron used can make a profound difference on the normal state
transport properties.

\begin{figure}[t]
\begin{center}
\includegraphics[angle=0,width=140mm]{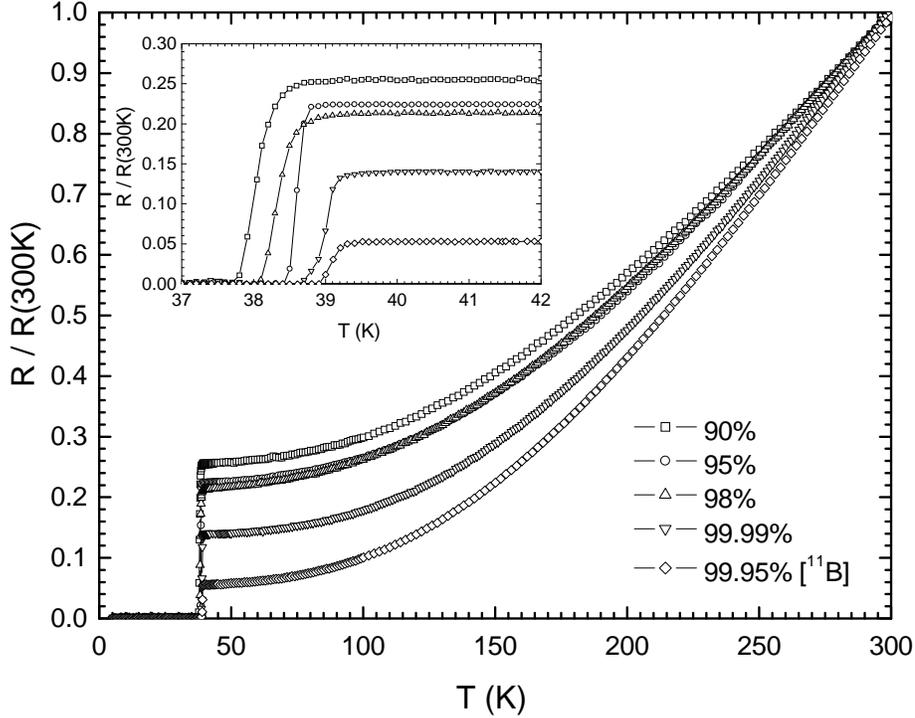}
\small\caption{\label{fig2} Variation of the zero-field
resistivity in the $5 K$ to $300 K$ range for MgB$_{2}$ pellets
with different boron purities. Inset shows a closeup at $T_{c}$.}
\end{center}
\end{figure}

In Fig. 3 the same resistance data is plotted, but instead of
simply normalizing the data at room temperature the data is
normalized to the temperature derivative at room temperature. This
is done to see if the resistance curves differ only by a
temperature independent residual resistivity term: i.e. this
normalization is based upon the assumption that the slope of the
temperature dependent resistivity at room temperature should be
dominated by phonon scattering and therefore be the same for each
of these samples. As can be seen this seems to be the case, at
least to the first order. By using higher purity boron we are able
to diminish the additive, residual resistance by a factor of
approximately five.

\begin{figure}[htb]
\begin{center}
\includegraphics[angle=0,width=140mm]{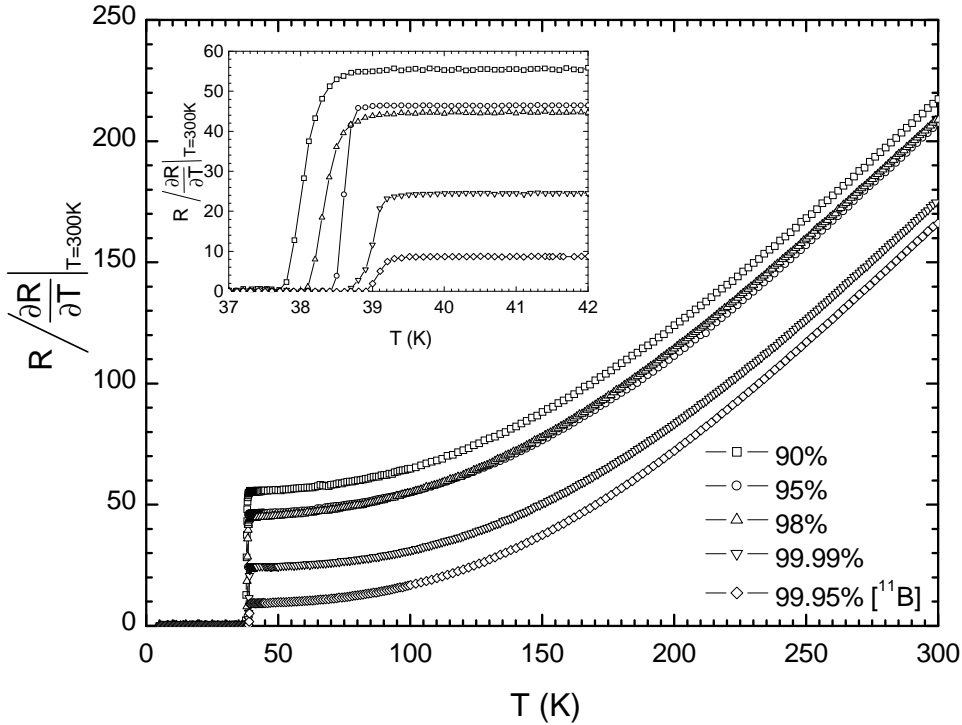}
\small\caption{\label{fig3} Resistivity curves normalized by
temperature derivative at room temperature, for different boron
purities.}
\end{center}
\end{figure}

The insets to figs. 2 and 3 also indicate that there is a
monotonic improvement in $T_{c}$ as the boron purity (or $RRR$
value) is increased. $T_{c}$ values vary from just below $38 K$ to
just above $39 K$ depending upon which boron is used.

Based upon these results we choose the isotopically pure $^{11}$B
for the further study of the effects of Mg stoichiometry on
MgB$_{2}$ pellet samples. But before we proceed to the next
section it is worth noting that one of the difficulties associated
with the samples made by other research groups may well be due to
the use of boron with less than the highest purity. In addition,
to our knowledge very few other groups have been using the
Eagle-Picher isotopically pure boron in the samples for electrical
transport measurements.

\section{Effects of Magnesium Stoichiometry}
\label{sec:stoichiometry}

\begin{figure}[b]
\begin{center}
\includegraphics[angle=0,width=136mm]{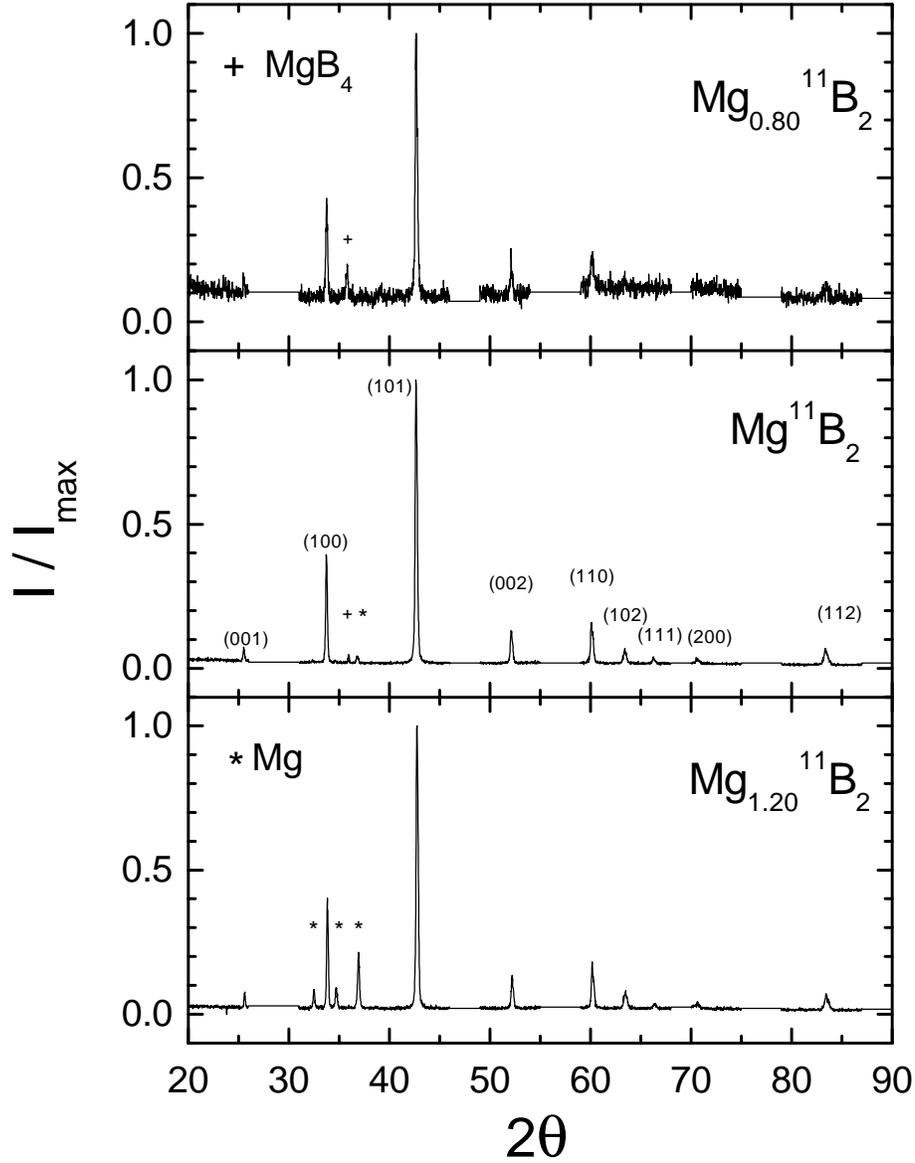}
\small\caption{\label{fig4} X-ray spectra for three different
nominal compositions of Mg$_{x}$$^{11}$B$_{2}$ for $x =$ 0.8, 1.0,
1.2. The symbols $\star$ indicates Mg peaks and $+$ indicates
MgB$_{4}$ peaks.}
\end{center}
\end{figure}

In order to study the effect of magnesium stoichiometry on the
transport properties of Mg$^{11}$B$_{2}$ a series of
Mg$_{x}$$^{11}$B$_{2}$ ($0.8 \leq x \leq 1.2$) samples were
synthesized. Figure 4 presents powder X-ray diffraction patterns
for the extreme members of the series (top and bottom panels) as
well as for the stoichiometric Mg$^{11}$B$_{2}$ (middle panel). In
all cases the lines associated with the Mg$^{11}$B$_{2}$ phase are
present. For the Mg$_{0.8}$$^{11}$B$_{2}$ sample there is a weak
line seen at $2\theta = 35.8^{\circ}$ that is associated with
MgB$_{4}$ (marked with a +). This is consistent with the fact that
there was insufficient Mg present to form single phase
Mg$^{11}$B$_{2}$. For the Mg$_{1.2}$$^{11}$B$_{2}$ sample there
are very strong diffraction lines associated with Mg (marked with
*). This too is consistent with the stoichiometry of the sample:
Mg$^{11}$B$_{2}$ is the most Mg-rich member of the binary phase
diagram, therefore any excess Mg will show up as unreacted Mg. The
X-ray diffraction pattern for the stoichiometric Mg$^{11}$B$_{2}$
shows much smaller peaks associated with a small amount of both
MgB$_{4}$ and Mg phases. This pattern is different from the one
shown in Fig. 2 in that this sample was reacted for three hours
whereas the sample used in Fig. 2 was reacted for four hours.
Given that all of the samples used for the Mg-stoichiometry study
were reacted for 3 hours it is appropriate to show this powder
diffraction set along with the other members of the series. It
should be noted that there is continuous change in the nature of
the second phases in the samples. For Mg deficient samples there
is only MgB$_{4}$ as a second phase. For the stoichiometric
Mg$^{11}$B$_{2}$ samples there are either no second phases or very
small amounts of both MgB$_{4}$ and Mg (depending upon reaction
times), and for the excess Mg samples there is no MgB$_{4}$, but
clear evidence of excess Mg.

\begin{figure}[b]
\begin{center}
\includegraphics[angle=0,width=130mm]{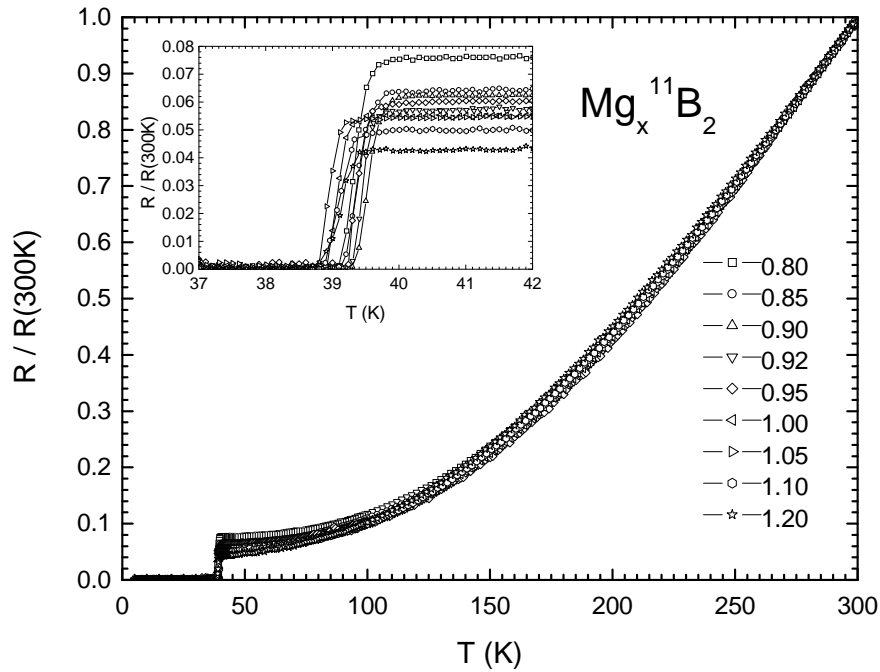}
\small\caption{\label{fig5} Temperature dependence of the
normalized resistivity for pellets with nominal composition
Mg$_{x}$$^{11}$B$_{2}$ ($0.8 < x < 1.2$).}
\end{center}
\end{figure}

Figure 5 presents normalized resistance data for eight different
Mg$_{x}$$^{11}$B$_{2}$ pellets. In each case the curve plotted is
the average of three or more samples cut from the same pellet.
There is far less variation between the different pellets in this
case than there was for the case of boron purity (Fig. 2). Figure
6 plots the $RRR$ values for each of the individual samples (shown
as the smaller symbols) as well as the $RRR$ of the average curve.
As can be seen the $RRR$ values increase slowly from $\sim 14$ for
Mg$_{0.8}$$^{11}$B$_{2}$ to $\sim 18$ for Mg$^{11}$B$_{2}$. This
is followed by a clear increase in $RRR$ values for excess Mg,
with Mg$_{1.2}$$^{11}$B$_{2}$ having an $RRR$ value of $\sim 24$.
The important point to note is that even for the most Mg deficient
sample the lowest measured $RRR$ value is significantly greater
than 10. At no point in this series we find samples with $RRR$
values of 3, 6, or 10, even when a clear MgB$_{4}$ second phase is
present. For samples ranging from Mg$_{0.9}$$^{11}$B$_{2}$ to
Mg$_{1.1}$$^{11}$B$_{2}$ (dotted box in figure 6) the average
$RRR$ values cluster around $RRR = 18 \pm 3$. These data indicate
that for sintered pellets $RRR$ values of 18 can be associated
with stoichiometric Mg$^{11}$B$_{2}$.

\begin{figure}[b]
\begin{center}
\includegraphics[angle=0,width=130mm]{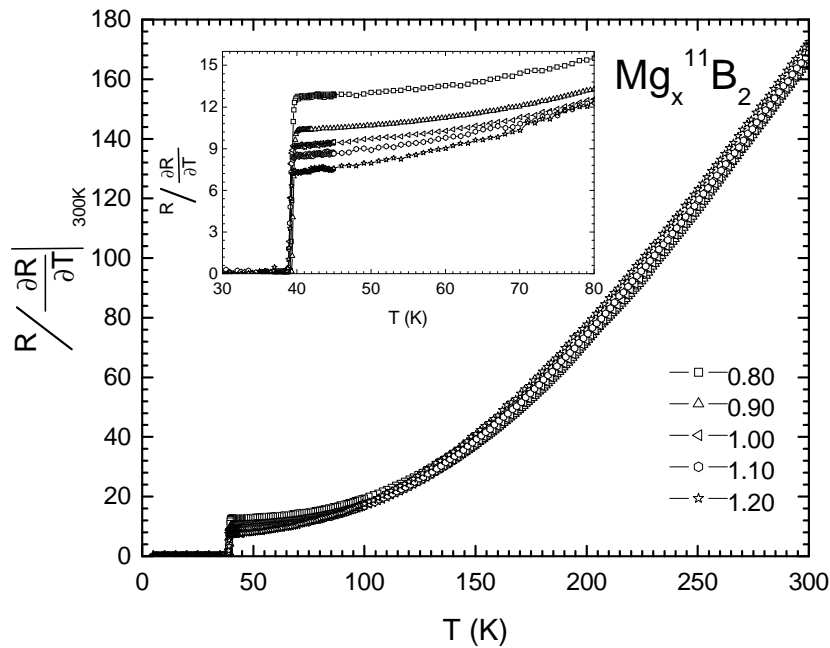}
\small\caption{\label{fig6} Residual resistance ratio of
Mg$_{x}$$^{11}$B$_{2}$ ($0.8 < x < 1.2$). The open symbols
represent different pieces selected from the same batch. The solid
symbols are the average. The dotted box delimits the small
variation ($x\pm 0.1$) with respect to the stoichiometric
compound.}
\end{center}
\end{figure}

Whereas the effects of excess Mg are relatively minor in these
samples (given their low intrinsic resistivities) these effects
can be clearly seen. In addition to the increase in the $RRR$
value there is a change in the form of the temperature dependence
of the resistivity. This can be best seen in Fig. 7 in which the
resistance data have been normalized its room temperature slope.
The data for all $x$ values less than 1.0 are similar and collapse
into a single curve. On the other hand the resistance data for the
$x=1.1$ and $x=1.2$ are qualitatively different. They start out
with somewhat higher normalized resistance data than the
stoichiometric sample and then below $100 K$ cross below the
stoichiometric sample (Fig. 7 inset). This is very likely due to
the increasing effects of having Mg in parallel (and series) with
the MgB$_{2}$ grains. As can be seen in Fig. 7 this effect becomes
larger as the amount of excess Mg is increased. This deviation
from the MgB$_{2}$\ resistance curve may actually serve as a
diagnostic for the detection of excess Mg.

\begin{figure}[htb]
\begin{center}
\includegraphics[angle=0,width=140mm]{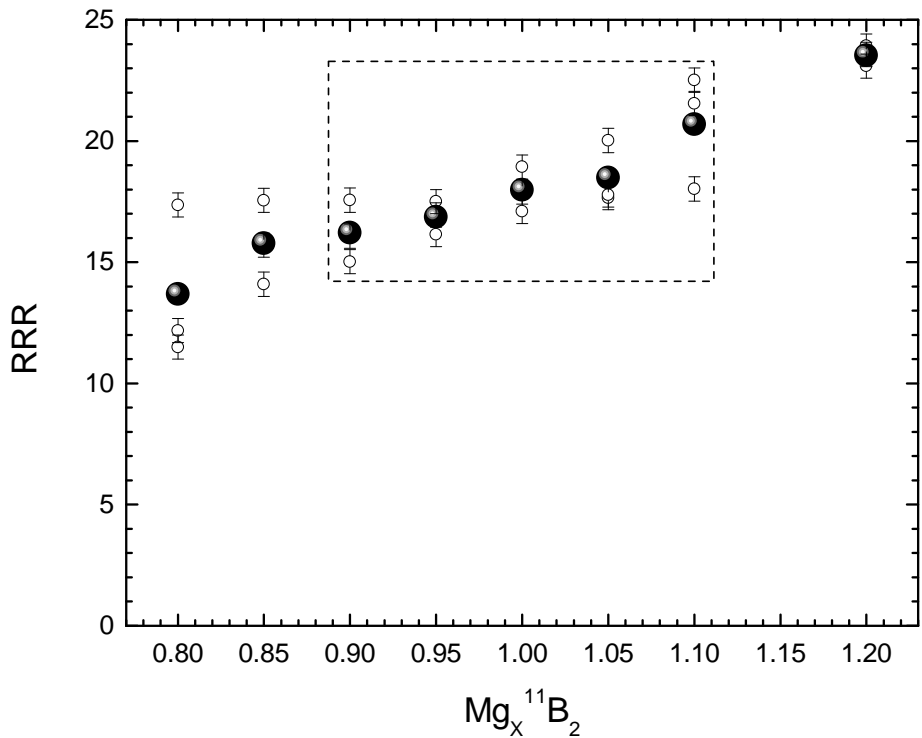}
\small\caption{\label{fig7} Resistivity curves normalized by
temperature derivative at room temperature, for
Mg$_{x}$$^{11}$B$_{2}$ ($x =$ 0.8, 0.9, 1.0, 1.1 and 1.2).}
\end{center}
\end{figure}

\section{MgB$_{2}$ wire segments: Effects of diameter,
etching and distilling} \label{sec:wire}

MgB$_{2}$ can also be synthesized in the form of wire segments
\cite{canfield01}. The starting material is boron filament that
has a small tungsten-boride core. Upon exposure to Mg vapor the
boron filament is transformed into MgB$_{2}$ wire segments, a
process that is accompanied by an expansion of the wire diameter.
Table 2 presents data on the initial and final diameters of the
boron fiber and MgB$_{2}$ wire segments used for this study. There
is a clear expansion associated with the transformation of the
boron into MgB$_{2}$. The average increase in the diameter
associated with this reaction is $\sim 1.4$ times. There is some
uncertainty associated with this number due to the fact that once
the MgB$_{2}$ is formed the wire segments have a somewhat
irregular surface as well as variation of the diameter along the
length of a segment. The tungsten boride core does not manifest a
noticeable change in diameter during this process. The size of the
tungsten boride core of the wire segments is listed on the far
side of Table 2. For the rest of this paper the wire segments will
be identified by the diameter of the initial boron filament used
to create them.

\begin{center}
\small Table 2: Main properties of MgB$_{2}$ wires and resistivity
at $300 K$ and $RRR$ for as-grown, etched and distilled MgB$_{2}$
filaments with different diameters.(Note: Samples by the diameter
of initial boron filaments used.)
\begin{tabular}{|c|c|c|c|}\hline
\multicolumn{2}{|c|}{ \begin{tabular}{c}
     \bf{Approximate Diameter [$\pm 10 \mu m$]} \\\hline
      \begin{tabular}{c|c}
        \bf{Initial (Boron)} & \bf{Final (MgB$_{2}$)}\\
    \end{tabular}  \\
    \end{tabular}}                      &\multicolumn{1}{c|}{ \bf{\begin{tabular}{c} Approximate\\ Expansion\\   \end{tabular} }}         & \multicolumn{1}{c|}{ \begin{tabular}{c} \bf{WB$_{x}$} \\
                                                                                                                             \bf{Approximate} \\
                                                                                                                                            \bf{Diameter [$\mu m$]} \\
                                                                                                                                              \end{tabular} }
\\\hline
 \multicolumn{2}{|c|}{ \begin{tabular}{c|c}
        100 & 140\\
       \end{tabular}}                      &\multicolumn{1}{c|}{1.4 }         & \multicolumn{1}{c|}{ 15}  \\\hline
\multicolumn{2}{|c|}{ \begin{tabular}{c|c}
        140 & 190 \\
       \end{tabular}}                                      &  1.5            & 15  \\\hline
\multicolumn{2}{|c|}{ \begin{tabular}{c|c}
         190 & 290  \\
       \end{tabular}}                                    & 1.5             & 15 \\\hline
\multicolumn{2}{|c|}{ \begin{tabular}{c|c}
         300 & 370  \\
       \end{tabular}}                                    & 1.3             & 20 \\\hline
\hline\hline

\multicolumn{2}{|c|}{ \bf{Sample Wires}}& $\mathbf{\rho(300K)[\mu
\Omega cm]}$ & $\mathbf{RRR}$ \\\hline
   & 100 & 13.0 & 18 \\
  AS & 140 & 10.8 & 45 \\
  GROWN & 190 & 8.3 & 25 \\
  & 300 & 14.4 & 36 \\\hline
  & 100 & 17.5 & 39 \\
  ETCHED& 140 & 15.8 & 30 \\
  & 190 & 9.1 & 45 \\
  & 300 & 16.3 & 28 \\\hline
  DISTILLED & 300 & 9.3 & 35 \\ \hline
\end{tabular}
\end{center}

Given that the wire segments are synthesized in a Mg rich vapor
(the nominal stoichiometry is Mg$_{3}$B), there is concern that
small amounts of excess Mg vapor condense onto the surface of the
MgB$_{2}$ wire segments during the cooling process. This could
lead to contributions to the temperature dependent resistivity
from metallic Mg. In addition there is the possibility that the
tungsten boride core may act as a low resistance resistor in
parallel with the MgB$_{2}$. Measurements of the transport
properties of MgB$_{2}$ wire segments of varying diameters will
allow us to examine, and ultimately discount, both of these
concerns.

If there were to be a significant contribution from metallic Mg on
the surface of the wire segments, and if it is to be assumed that
the metallic Mg has a lower resistivity than the intrinsic
resistivity of the MgB$_{2}$ (an assumption that is supported
experimentally by our earlier data on the
Mg$_{1.2}$$^{11}$B$_{2}$), then the effect of this excess Mg would
scale with the surface area to volume ratio of the sample: i.e.
there would be a substantially larger effect seen for the smaller
diameter wires than for the larger diameter wires. In a similar
manner the potential effect of the tungsten boride core would
scale with the square of the ratio of the tungsten boride diameter
to the MgB$_{2}$ wire diameter. Given that the tungsten boride
diameter remains between 15 and $20 \mu m$ over the whole series
and that the MgB$_{2}$ diameter increased from $\sim 140 \mu m$ to
$\sim 370 \mu m$, the potential effect that the tungsten boride
would have would be largest in the smaller diameter wires and
smaller in the larger diameter wires.

The temperature dependencies of the normalized resistivity of
MgB$_{2}$ wire segments are shown in Fig. 8a. All four diameter
wires have similar temperature dependencies, but manifest somewhat
different $RRR$ values. The inset to Fig. 8a shows the low
temperature behavior near $T_{c}$. From this plot it becomes clear
that there is no apparent effect of metallic Mg or tungsten boride
on the resistivity. The $RRR$ values for the 100, 140, 190 and
$300 \mu m$ wire segments are 18, 45, 41 and 25 respectively. The
highest $RRR$ value is for $140 \mu m$ and the lowest value is for
$100 \mu m$. To first order, there appears to be little or no
correlation between the wire diameter and the $RRR$ values. If any
trend is to be extracted it is that $RRR$ values are generally
higher for the larger diameter wires, a trend that contradicts the
assumption that metallic Mg or tungsten boride are affecting the
resistivity measurements.

\begin{figure}[htb]
\begin{center}
\includegraphics[angle=0,width=140mm]{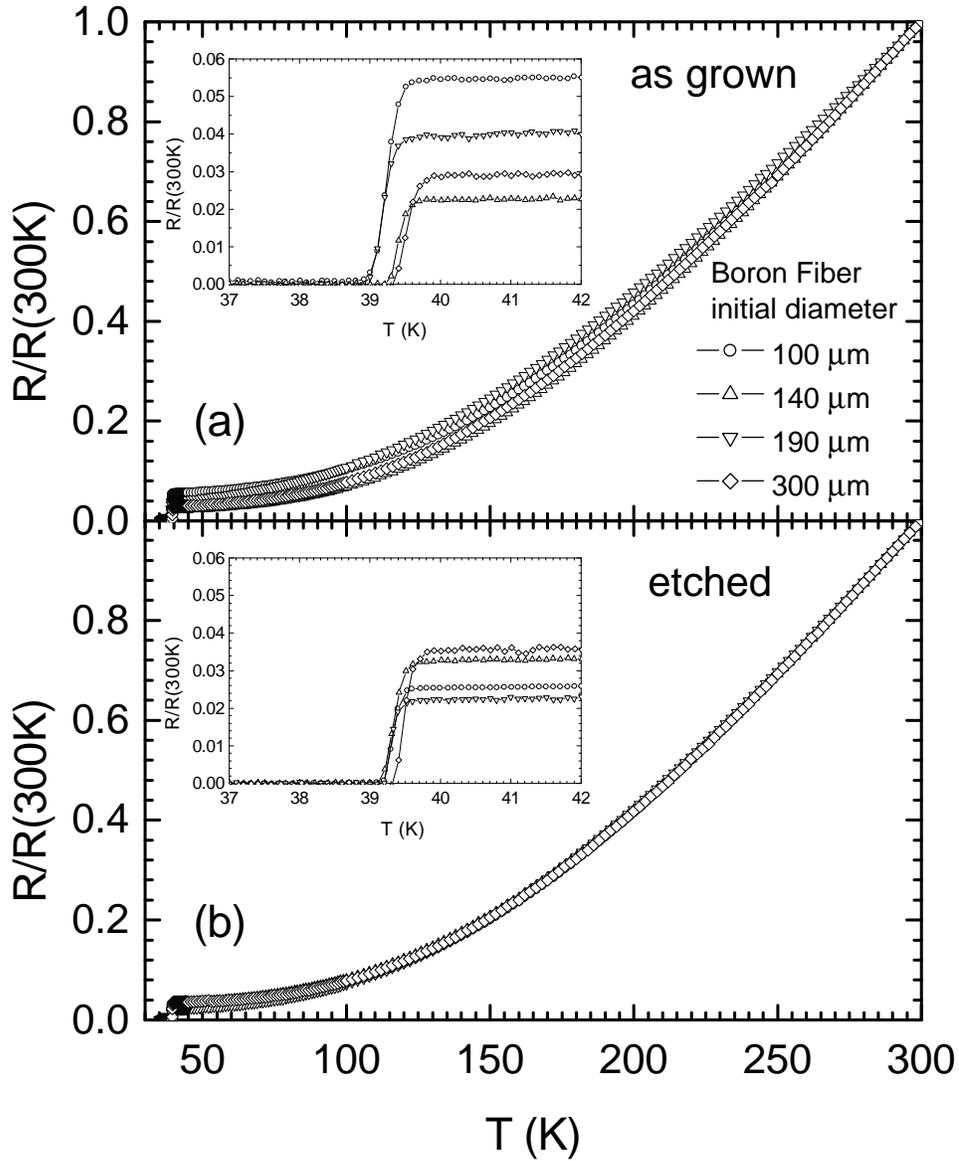}
\small\caption{\label{fig8} Resistivity curves of MgB$_{2}$
filaments with four boron fiber initial diameters (100, 140, 190
and $300 \mu m$): (a) for as-grown wires and (b) after etched in
ethyl alcohol with 5\% of HCl.}
\end{center}
\end{figure}

In order to further examine the possible effects of metallic Mg on
the transport properties of the wires we etched the as-grown wire
segments in a solution of 5\% HCl in ethyl alcohol for 5 minutes.
This lead to an apparent removal of any Mg coating on the wire
surface. The temperature dependence of the normalized resistance
of these etched samples are shown in Fig. 8b. In this case the
values of $RRR$ are increased.  If the excess Mg in wires were
acting as a parallel resistance in the sense of shorting
MgB$_{2}$, we would expect that with its removal we would find
smaller $RRR$ values, not larger.

In the distillation process, the MgB$_{2}$ as-grown wire with $300
\mu m$ boron fiber initial diameter, which has $RRR = 36$ was
heated to $600^{\circ}C$ for 12 hours and submitted to a
continuously pumped high vacuum for removal of any excess Mg. This
process practically did not alter the value of $RRR \sim $ 35.
This is strong evidence that the high values of $RRR$ which we
obtained for MgB$_{2}$ wires are intrinsic and not an influence of
excess Mg.

\begin{figure}[htb]
\begin{center}
\includegraphics[angle=0,width=140mm]{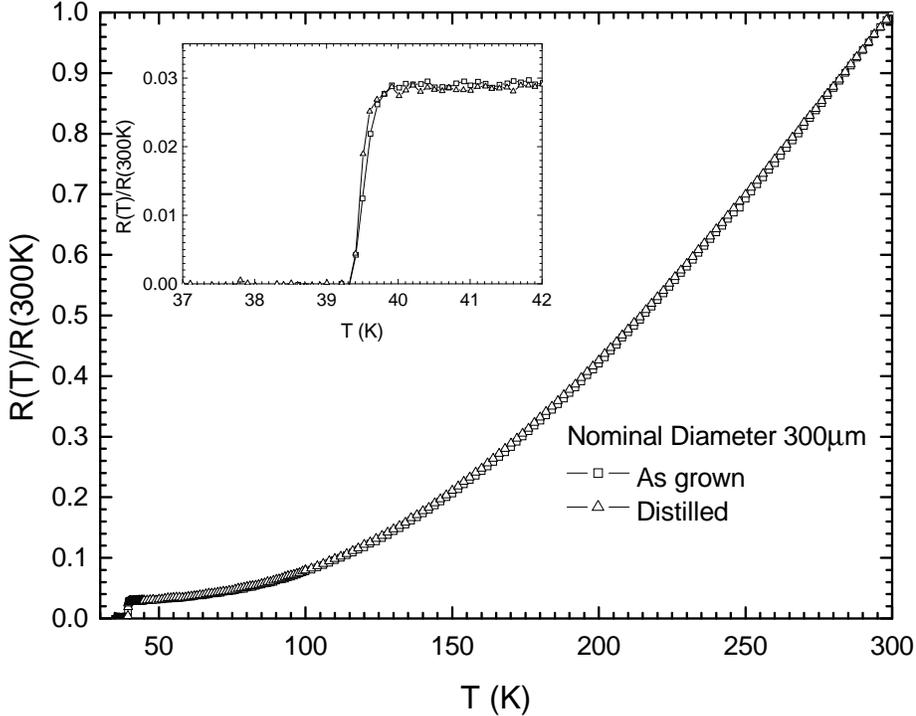}
\small\caption{\label{fig9} Resistivity curves before and after
distillation process in MgB$_{2}$ wire with $300 \mu m$ boron
fiber initial diameter.}
\end{center}
\end{figure}

Table 2 presents our estimates of the room temperature resistivity
for each wire sample. The average room temperature resistivity for
these nine samples is roughly 13 $\mu \Omega cm$. It should be
noted that the acid etching does increase the measured resistivity
of the wires. We believe that this is due to two effects: i) a
reduction of cross sectional area that has not been accounted for
and ii) the creation of small cracks in the wire. In both cases
the will change the effective geometry of the sample.

A final point about all of the wire samples is worth noting: all
of the measured $RRR$ values are comparable to or better than
those found for the stoichiometric $^{11}$B pellet samples. These
boron fiber are made using a boron fluoride intermediate step and
are reported to be 99.999\% pure. This again points out that the
purity (and probably the purification process) of the boron may be
a critical variable.

\section{Conclusion} \label{sec:conclusion}

In summary, through the synthesis of various pellets of MgB$_{2}$
with different types of boron we found values of $RRR$ from 4 to
20, which covers almost all values found in literature. To obtain
high values of $RRR$, high purity reagents are necessary. With the
isotopically pure boron we obtained the highest $RRR \sim$ 20 for
the stoichiometric compound. We also investigated
Mg$_{x}$$^{11}$B$_{2}$ samples with 0.8 $< x <$ 1.2. These have
shown that from the most Mg deficient samples we observe
inclusions of the MgB$_{4}$ phase, and no evidence of Mg. For
samples with excess Mg we do not observe any MgB$_{4}$. For the
range Mg$_{0.8}$$^{11}$B$_{2}$ up to Mg$_{1.2}$$^{11}$B$_{2}$ we
found average values of $RRR$ between 14 and 24. For smaller
variations in stoichiometry ($x\pm 0.1$) $RRR =  18 \pm 3$. In
addition our study of MgB$_{2}$ wires as function of diameter is
consistent with pellet results and inconsistent with either Mg or
tungsten boride core acting as resistor in parallel with MgB$_{2}$
filaments. All of our data are point to conclusion that high $RRR$
($\geq 20$) and low $\rho_{0}$ ($\leq 0.4 \mu \Omega cm$) are
intrinsic materials properties associated with high purity
MgB$_{2}$.

{\bf Acknowledgements}

Ames Laboratory is operated for the US Department of Energy by
Iowa State University under Contract No. W-7405-Eng-82. This work
was supported by the Director for Energy Research, Office of Basic
Energy Sciences and the National Science Foundation under grant
No. DMR-9624778. The authors would like to thank D. K. Finnemore,
M. A. Avila and N. Anderson for helpful assistance and many
fruitful discussions.

\end{document}